\documentclass[fleqn,twoside]{article}
\usepackage{espcrc2}

\usepackage{graphicx}
\usepackage[figuresright]{rotating}

\newcommand{\half}{ {\scriptstyle \frac{1}{2} } }

\newcommand\be{\begin{equation}}
\newcommand\ee{\end{equation}}
\newcommand\bea{\begin{eqnarray}}
\newcommand\eea{\end{eqnarray}}
\newcommand{\bdm}{\begin{displaymath}}
\newcommand{\edm}{\end{displaymath}}
\newcommand\nn{ \nonumber\\}
\newcommand\n{ \nonumber}

\newcommand{\tbox}[1]{\quad \mbox{#1} \quad}

\newcommand{\dd}[1]{\partial_{#1}}
\def\dd{\partial}
\newcommand{\Tr}{\mbox{Tr}}

\newcommand{\AmS}{{\protect\the\textfont2 
  A\kern-.1667em\lower.5ex\hbox{M}\kern-.125emS}}

\title{Chiral Extension of Lattice Gauge Theory}

\author{Richard C. Brower\address[MCSD]{ 
Physics Department\\Boston University\\
Boston, MA 02215, USA .}}
       
\begin{document}

\begin{abstract}
Two approaches are presented to coupling explicit Goldstone modes to
$N_f$ flavors of massless quarks preserving exact $SU(N_f) \times
SU(N_f)$ chiral symmetry on the lattice. The first approach is a
generalization a chiral extension to QCD (aka XQCD) proposed by
Brower, Shen and Tan consistent with the
Ginsparg-Wilson relation.  The second approach based on the Callan,
Coleman, Wess and Zumino coset construction has a real determinant at
zero quark axial coupling, $g_A = 0$. 
\vspace{1pc}
\end{abstract}

\maketitle

\section{INTRODUCTION}
A persistent difficulty with the standard lattice approach to low
energy hadronic processes is the difficulty with small eigenvalues in
the limit of small quark mass, which through the Banks-Cashir formula
are responsible for chiral symmetry breaking. Since these eigenvalues
are represented faithfully by bosonic matrix models, one might hope
that they could somehow be ``bosonized''. In 1994, Brower, Shen and
Tan~\cite{Brower} introduced a new methods called ``chirally extended
QCD'' (or $XQCD$) in this spirit. The lattice action is modified by
adding explicit fields for the Goldstone modes that has the effect of
replacing the low eigenvalues by a constituent quark mass $M_Q$ {\bf
without} explicitly breaking $SU_L(N_f) \times SU_R(N_f)$ chiral
symmetry.

Here we extend this method to incorporate chiral fermions obeying the
Wilson-Ginsparg relation. In this way we believe the continuum limit
for XQCD on the lattice will approach the universal fixed point for
QCD without fine tuning. These methods are of general interest for the
 study of non-perturbative models of chiral symmetry and Higgs
symmetry breaking phenomena on the lattice. 

\section{LATTICE REALIZATIONS}

In continuum notation, the Lagrangians we wish to put on the lattice have the
general form,
\be
{\cal L}_{\chi QCD} = {\cal L}_{YM} + {\cal L}_{\chi} + {\cal L}_{F} \; ,
\ee
where the Yang Mills theory  and
non-linear chiral Lagrangian for the Goldstone modes
\be
{\cal L}_{YM} = \frac{1}{4}Tr[F^2_{\mu\nu}] \tbox{and} {\cal L}_\chi = \frac{F^2_\pi}{4} \Tr[\dd_\mu\Sigma^2] + \cdots
\ee
respectively are coupled through Yukawa interactions in the Fermion action,
\be
{\cal L}_{F} = \bar \psi\gamma_\mu (\partial_\mu -
i g_{YM} A_\mu) \psi +  M_Q \bar \psi
\Sigma^{(5)}\psi  \; .
\ee
Here $\Sigma$ is an element of $SU(N_f)$ and we define
\be
\Sigma^{(5)} \equiv \frac{1 +\gamma_5}{2}\Sigma + \frac{1 - \gamma_5}{2} \Sigma^\dagger  \equiv \exp[ 2 i  \gamma_5 \vec \pi \cdot \vec \tau/F_\pi]   \; .
\ee 
One can place this on a lattice by introducing a naive Fermions action
\be
S_{F}  =   \half \bar \Psi_x[\gamma_\mu U_\mu(x)
\delta_{x+1,y}  - \gamma_\mu U^\dagger(x)\delta_{x,y+1}] \Psi_y  
\ee
plus a Wilson-Yukawa term that removes doublers and
respects chiral invariance:
\bea
&&\bar \Psi_x [ \frac{r}{2}   (\delta_{xy} -  U_\mu(x)\delta_{x+1,y} - U^\dagger(x)
\delta_{x,y+1})\nn
&&\qquad(\Sigma^{(5)}_x + \Sigma^{(5)}_y)  +  (M_Q  \Sigma^{(5)}_x   + m_q) \delta_{xy} ] \Psi_y  \; .\n
\eea
The only term that explicitly  breaks
$S(N_f) \times SU(N_f)$ symmetry is the quark mass term,
$m_q \bar \Psi_x \Psi_x$. Still if there is a continuum limit
which decouples the scaler fields, a delicate fine tuning must
be required to adjust the renormalized quark mass  to zero.
To avoid this problem we now consider coupling to Ginsparg-Wilson fermions.

\subsection{Overlap Fermions}

With the standard overlap operator, $D_0 = 1 +  \gamma_5 \epsilon(H_W)$,
the Ginsparg-Wilson relation can be express in two
alternative (asymmetric) forms,
\be
\gamma_5 D_0 + D_0 \widehat \gamma_5 = 0
\tbox{and} \widehat \gamma'_5 D_0 + D_0  \gamma_5 = 0
\ee
with $\widehat \gamma_5 \equiv \gamma_5 (2 - D_0)$ and $\widehat
\gamma'_5 \equiv (2 - D_0)\gamma_5 $ respectively.  Thus there are two
different realizations of lattice chiral symmetry with non-local
projectors $ 1 \pm \widehat \gamma_5$ or $1 \pm \widehat \gamma'_5$
acting on the ``ket's'' or ``bra's''respectively. This leads to two
ways to embed the chiral extension into the Ginsparg-Wilson
relation. We will show that they are both equivalent up to a field
redefinition and indeed both can be derived from a single form of the
Domain Wall action.

The construction is  straight forward. In the first case the
chiral transformations become,
\bea
\Psi &\rightarrow& \exp[i \widehat \gamma_5\vec  \theta_A \cdot \vec \lambda + i \vec \theta_V  \cdot \vec \lambda] \Psi \nn
\bar \Psi &\rightarrow& \bar \Psi  \exp[i \gamma_5 \vec \theta_A  \cdot \vec \lambda - i \vec\theta_V  \cdot \vec\lambda]
\eea
The invariant Fermion term is 
$$
\bar \Psi_x D_{0,x,y} \Psi_y + M_Q [ \bar\Psi_{L,x} \Sigma_x \Psi_{R,x} + \bar\Psi_{R,x}\Sigma^\dagger_x \Psi_{L,x}]
$$
or a new XQCD operator,
\be 
 D(U,\Sigma) = D_0(U) + M_Q \Sigma^{(5)} (1 - D_0(U))  \; .
\ee
Expanding $\Sigma^{(5)} = \widehat \sigma(x) + i \gamma_5 \vec \tau \cdot
\vec{\widehat \pi}(x)$ for $N_f = 2$, we have rather familiar looking form,
$$
D(U,\Sigma)  = D_0 + M_Q \widehat \sigma \frac{1 + \gamma_5 \widehat \gamma_5}{2} + i M_Q \vec \tau \cdot \vec{\widehat \pi}(x) \frac{\gamma_5 + \widehat \gamma_5}{2} \; .  
$$
The alternative form
is
\be
\widetilde D(U,\Sigma) = D_0 + M_Q  (1 - D_0) \Sigma^{(5)} 
\ee
One readily shows that the measure preserving field redefinition
$\Psi_x \rightarrow \Psi'_x = (1 - D_0) \Psi_x$, and $\bar \Psi_x
\rightarrow \bar \Psi'_x = \bar \Psi_x (1 - D_0)^{-1} $ converts one
form into the other.

\subsection{Coset Construction}

The coset construction for the chiral breaking of a symmetry group
G to a subgroup H, fixes the frame Fermion field, $\Psi_x = \gamma(x)Q_x $,
by choosing a canonical element $\gamma(x)$ in the coset G/H. In our present case, the natural choice is 
\bea
\gamma(x) &=& \xi^{(5)\dagger} \equiv \exp[- i \gamma_5 \vec \pi \cdot \vec \tau/F_\pi] \nn  
&=& \half(1+\gamma_5) \xi^\dagger + \half(1-\gamma_5) \xi  \; .
\eea 
Now chiral symmetry requires  a rotation of the new quark field $Q_x$ to re-establishes the frame: $Q_x \rightarrow u_x Q_x \tbox{,} \bar Q_x \rightarrow  \bar Q_x u^\dagger(x)$, when $\xi_x \rightarrow L \xi_x u^\dagger(x) = u(x) \xi_x R^\dagger$.

There are a variety of ways to proceed to construction
invariant lattice Lagragians. One nice way is to
introduce bilocal lattice
``currents'':
$j_R =  \xi^{(5)}_x   \xi^{\dagger (5)}_y = V_{xy} + i \gamma_5 A_{xy}$
and $j_L =  \xi^{\dagger (5)}_x   \xi^{(5)}_y = V_{xy} - i \gamma_5 A_{xy}$
where
$V_{xy} =  \frac{1}{2} ( \xi^\dagger_x \xi_y + \xi_x \xi^\dagger_y )
\tbox{and} A_{xy} = \frac{1}{2i}( \xi^\dagger_x \xi_y -  \xi_x \xi^\dagger_y) 
$
transform as adjoint flavor tensors: $ V_{xy} \rightarrow u_x V_{xy} u_y^\dagger
$, $ A_{xy} \rightarrow u_x A_{xy} u_y^\dagger $. 
Now the Fermion operator takes the attractive from, $D^{coset}_{xy}(U,\xi) = D_{0xy}(V_{xy}+ i g_A  A_{xy})$, leading to a general lattice action,
\bea
S_{XQCD} &=& \!\!\!\! \frac{6}{g^2_0} Tr[U_{\mu\nu}] + \bar Q_x D^{coset}_{xy}(U,\xi) Q_y  \nn
&+ &  \!\!\!\! M_Q \bar Q_x Q_x + \half m_q \bar Q_x\xi^{\dagger (5)}_x \xi^{\dagger (5)}_xQ(x) \nn
&+&  \!\!\!\! \frac{F_\pi}{4} \sum_{x,\mu} Tr[A(x,x+\mu)A^\dagger(x,x+\mu)]  \n
\eea
where the kinetic term for the non-linear sigma models again 
is rewritten by $ Tr[A_{x,x+\mu} A^\dagger_{x,x+\mu}]\nn
=  \frac{1}{4} Tr[ (\Sigma_{x+\mu} -  \Sigma_x)(\Sigma^\dagger_{x+\mu} -  \Sigma^\dagger_x)] $.
\begin{figure}
\begin{center}
\includegraphics[width=2.5in,height=3.0in,angle=-90]{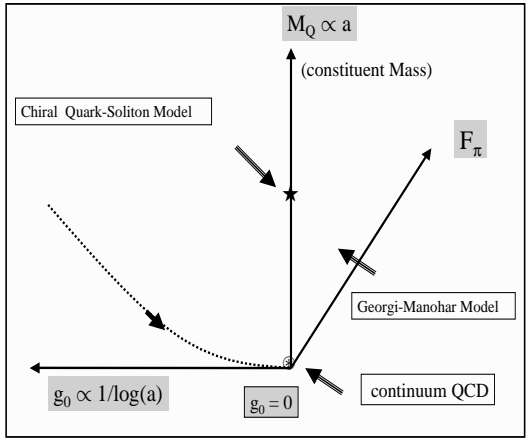}
\end{center}
\caption{Renormalization group flow at zero quark mass $m_q$ }
\label{fig:phasediag}
\end{figure}

\section{DISCUSSION}

More details will be
given in a forth coming publication by Berruto, Brower, Neff, Edwards,
Lim and Tan. However there are several remarks that should be made.   

As emphasized recently by Chandrasekharan, Pepe, Steffen and
Wiese~\cite{Chandrasekharan} the coset construction applies to
very wide class of effective chiral Lagragians. For example one can
replace the overlap operator $D_0$ above by the standard Wilson lattice
operator, still maintaining exact $SU(N_f) \times SU(N_f)$ up
to explicit quark mass terms.  Indeed the value of $g_A$ maybe
different in the naive kinetic term and the Wilson mass term and setting
them to $g_A = 1$ and $g_A = 0$ respectively gives precisely the
original Wilson-Yukawa construction with the field redefinition $Q_x
\rightarrow \xi^{(5)} \Psi_x$.  The most general clover improved
XQCD (i.e. using all 5-d operators) will be given elsewhere.

In general these action lead to a complex determinant. Indeed using the
gradient expansion for weak fields one can show that first contribution to
the phase linear in $g_A$ is proportional to 13-d operators such as:
$$
\frac{g_A g^3}{m_Q^6}\ \epsilon_{ijk}
\partial^2_\lambda\pi^i\partial_{\mu}\pi^j\partial_{\nu}\pi^k
d_{abc}F_{\mu\nu}^aF_{\rho\tau}^b\widetilde{F}_{\rho\tau}^c
\label{rg}
$$
This is consistent with general theorems for a vanishing phase with $N_c
\le 2$ or $N_f < 2$ or $d \le 3$ or $g_A = 0$, etc. 

Finally the general form of our Lagrangian for XQCD has 3 basic parameters:
the bare gauge coupling,$g_0$, the constituent quark mass $F_\pi$ and the
bare pion decay constant $F_\pi$ (see Fig.~\ref{fig:phasediag}). Special
limits afford interesting models. The Georgi-Manohar chiral quark model is
$g_0 \rightarrow 0$, which reproduces the non-relativistic quark model
results with $g_A \simeq 0.75$ and $M_Q \simeq 350 Mev$. The Chiral Soliton
Quark model,which is believed to have proton and Delta bound states is $F_\pi
\rightarrow 0$,$g_0 \rightarrow 0$.  These non-linear effective chiral
lattice actions will allow a non-perturbative investigation of the phase diagram
(Fig.~\ref{fig:phasediag}) to clarify chiral symmetry breaking mechanism in
general and the best way to approach the chiral limit of QCD in particular.

\end{document}